\documentclass[letter]{aa}  

\usepackage{graphicx}
\usepackage{natbib}
\bibpunct{(}{)}{;}{a}{}{,}
\usepackage{txfonts}
\graphicspath{{./Figures/}}
\begin{document}
\title{First astronomical unit scale image of the GW Ori triple
  system} 
\subtitle{Direct detection of a new stellar companion.}

   \author{J.-P. Berger\inst{1,2}
          \and
          J. D. Monnier\inst{3}
          \and
          R. Millan-Gabet\inst{4}
\and
S. Renard\inst{2}
\and
E. Pedretti\inst{5}
\and
W. Traub\inst{6}
\and
C. Bechet\inst{13}
\and
M. Benisty\inst{7}
\and
N. Carleton\inst{8}
\and
P. Haguenauer\inst{1}
\and
P. Kern\inst{2}
\and
P. Labeye\inst{9}
\and
F. Longa\inst{2}
\and
M. Lacasse\inst{8}
\and
F. Malbet\inst{2}
\and
K. Perraut\inst{2}
\and
S. Ragland\inst{10}
\and
P. Schloerb\inst{11}
\and
P. A. Schuller\inst{12}
\and
E. Thi\'ebaut\inst{13}
}

   \offprints{J.-P. Berger}

\institute{European Southern Observatory, Alonso de Cordova 3107 Vitacura, Casilla
19001, Santiago de Chile 19, Chile \\
              \email{jpberger@eso.org}
\and
IPAG, CNRS/UMR 5571,
     Universit\'e J. Fourier, BP-53, F-38041 Grenoble Cedex, France 
        \and
            University of Michigan, 941 Dennison Building, 500 Church
             Street, Ann Arbor, MI 48109-1090.e USA
         \and California Institute of Technology, 770 S. Wilson
         Ave. MS 100-22 Pasadena, CA 91125,
         USA
         \and University of St. Andrews, Scotland, UK
\and
JPL, California Institute of Technology, M/S 301-355, 4800 Oak Grove Drive, Pasadena, CA 91109, USA
\and
INAF-Osservatorio Astrofisico di Arcetri, Largo E. Fermi 5, 50125
Firenze, Italy
\and
Harvard Smithsonian Center for Astrophysics, 60 Garden Street,
Cambridge, MA 02138, USA
\and
CEA-DRT-LETI, 17 rue des Martyrs,
38054 Grenoble, France
\and
W. M. Keck Observatory, 65-1120 Mamalahoa Hwy, Kamuela, HI 96743, USA
\and
University of Massachusetts, Department of Astronomy, Amherst, MA 01003~4610, USA
\and
  IAS, CNRS/UMR 8617, Universit\'e
  Paris-Sud, 91405 Orsay, France
\and CRAL Observatoire de Lyon 9 avenue Charles André 69561 Saint
Genis Laval, France
}
 \date{Received: December 2010}

 
  \abstract
  {Young and close multiple systems are unique laboratories to probe
    the initial dynamical interactions
    between forming stellar systems and their dust and gas environment. Their study is a key
    building block to understanding the high frequency of main-sequence multiple systems. However, the number of detected
    spectroscopic young multiple systems that allow dynamical studies is
    limited. GW Orionis is one such system. It is one of the
    brightest young T Tauri stars and is surrounded by a massive disk.}
   {Our goal is to probe the GW Orionis multiplicity at angular scales
     at which we can spatially resolve the orbit.}
   {We used the IOTA/IONIC3 interferometer to probe the 
   environment of GW Orionis with an astronomical
   unit resolution in 2003, 2004, and 2005. By measuring squared
   visibilities and closure phases with a good UV coverage we carry
   out the first image reconstruction of GW Ori from
   infrared long-baseline interferometry.}
 {\emph{We obtained the first infrared image of a T Tauri multiple system with astronomical unit resolution.} We show that GW Orionis is a triple system,
   resolve for the first time the previously known inner pair
   (separation $\rho\sim$1.4~AU) and reveal a new more distant
   component (GW~Ori C) with a projected separation of $\sim$8~AU with direct
   evidence of motion.  Furthermore,
   the nearly equal (2:1) H-band flux ratio of the inner components
   suggests that either GW~Ori B is undergoing a preferential
   accretion event that increases its disk luminosity or that the
   estimate of the masses has to be revisited in favour
of a more equal mass-ratio system that is seen at lower inclination.}  {Accretion disk models of GW~Ori will need
   to be completely reconsidered because of this
   outer companion C and the unexpected brightness of companion B.}


   \maketitle
%

\section{Introduction}

\begin{table*}[t]
\centering
{\tiny
\begin{tabular}{|l|p{2cm}|p{2cm} | p{2cm} | p{3cm}|}
\hline
Array config. &  UT Date & Min. Base. (m) &
Max. Base. (m)&    Calibrators (diameters) \\
\hline
\hline
A35B15C10 & 2003-11-30 & 16.9 & 36.6 &HD42807 ($0.45 \pm 0.05$mas)   \\
A35B15C10 & 2003-12-01 & " & " & HD 42807 \\
\hline
\hline
A28B10C0 & 2004-12-12 & 7.9 & 29.9 & HD31966, HD42807  \\
A28B10C0 & 2004-12-14 & " & " & HD31966, HD42807      \\
A28B05C10 & 2004-12-15 & 10.3 & 28.6& HD31966, HD42807   \\
A28B05C10 & 2004-12-16 & " & " & HD31966, HD42807 \\
A28B05C10 & 2004-12-20 & " & " & HD31966, HD42807  \\
\hline
\hline
A35B15C0  & 2005-11-22 & 14.3 & 36.7 & HD31966 ($0.43 \pm 0.06$mas)  \\
\hline
\end{tabular}
}
\caption{\label{tab:logs} Epochs and array configurations of the observations. 
Adopted diameter  used for calibrator HD 42807 (G2V, V=6.44, H=5.01):
$0.45 \pm 0.05$ mas and for HD31966 (G5V, V=6,7, H=5.2): $0.43 \pm 0.06$
mas (from getcal/NexSci). Minimum and maximum projected baselines for each configuration
are given as an indication. }
\end{table*}

Stellar initial masses are controlled by the accretion of gas through
disks onto the star in the protostellar phase. For
a binary or multiple system this process is more complex because it
potentially involves the presence of circumstellar and circumbinary
gas reservoirs that control the mass distribution
\citep{Bate:1997,Ochi:2005}.  Understanding the origin of such a
distribution therefore requires the observation of multiple systems,
the characterisations of their masses, and of the nature of the accretion sources.

GW Orionis is a single-line spectroscopic binary Classical T Tauri
Star with an orbital period of 242 days located at
$\approx 400$ pc \citep{Mathieu:1991}\footnote{Distance confirmed by
  \citet{Sandstrom:2007,Menten:2007}}. The primary star's mass has
been estimated to be 2.5 M$_{\odot}$ and the secondary's as 0.5 M$_{\odot}$
with an expected separation of $\sim$1 AU (circular orbit of inclination
$\approx 27^{\circ}$). Evidence of a third companion was found
based on the analysis of radial velocity residuals (confirmed period
of 3850 days; Latham, priv. comm.).  The spectral
energy distribution (SED) of the unresolved complete system shows a
large infrared excess compared with a photosphere from a 5662 K
G5 star \citep{Cohen:1979} with a strong 10-microns silicate emission
feature. An important near-to-mid-infrared dip in the SED has been
interpreted as the signature of a circumbinary (CB) disk, whose inner
part is clearing up to 3.3 AU from the star system
\citep{Mathieu:1991}. \citet{Artymowicz:1994} have attempted to
simulate the interaction of the CB disk with the central system. Photometric time-dimming behaviour
led \citet{Shevchenko:1998} to conclude that eclipses resulted from
material around GW~Ori~B that covers the central star. 

An accurate model of the disk structure is not possible without good
knowledge of the stellar components of this system. We present here
the first long-baseline interferometer data on GW~Ori using the
IOTA interferometer and succesfully obtain the first
reconstructed image of a T Tauri system with astronomical-unit
resolution.


\section{Observations and data reduction}

\begin{figure*}[t]
  \centering
\begin{tabular}{cc}
  \includegraphics[width=0.30\textwidth]{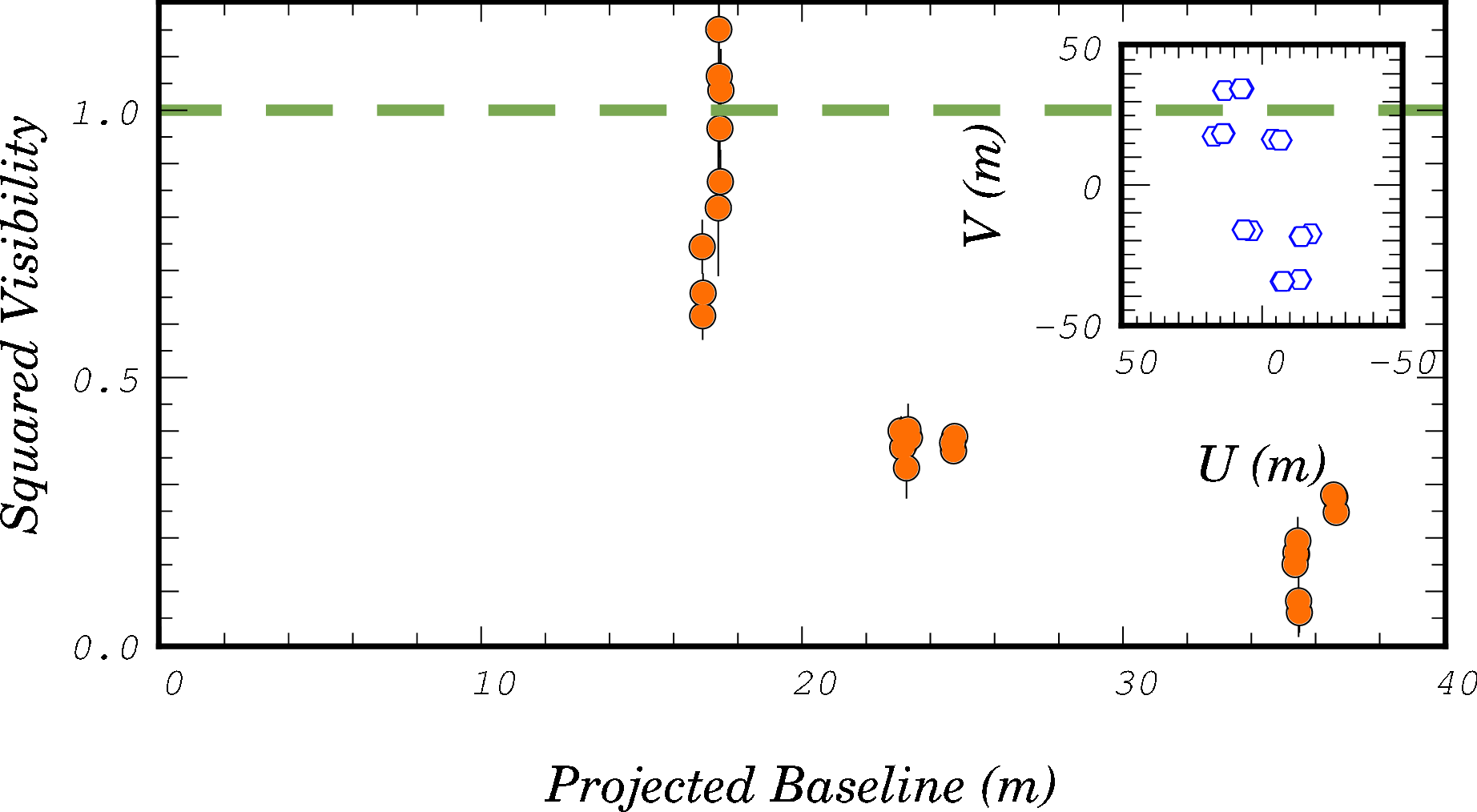}
&  \includegraphics[width=0.30\textwidth]{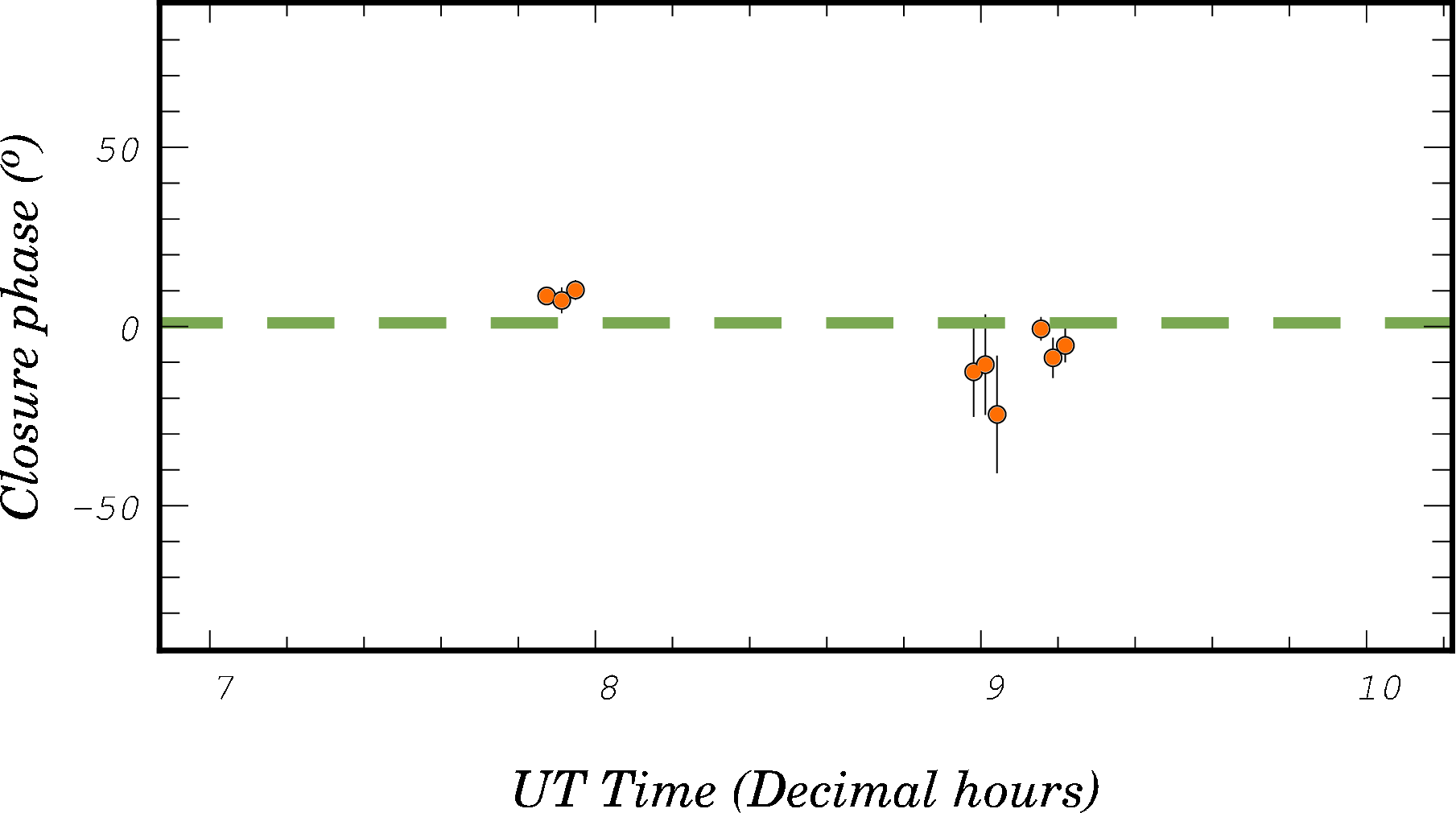}\\
  \includegraphics[width=0.30\textwidth]{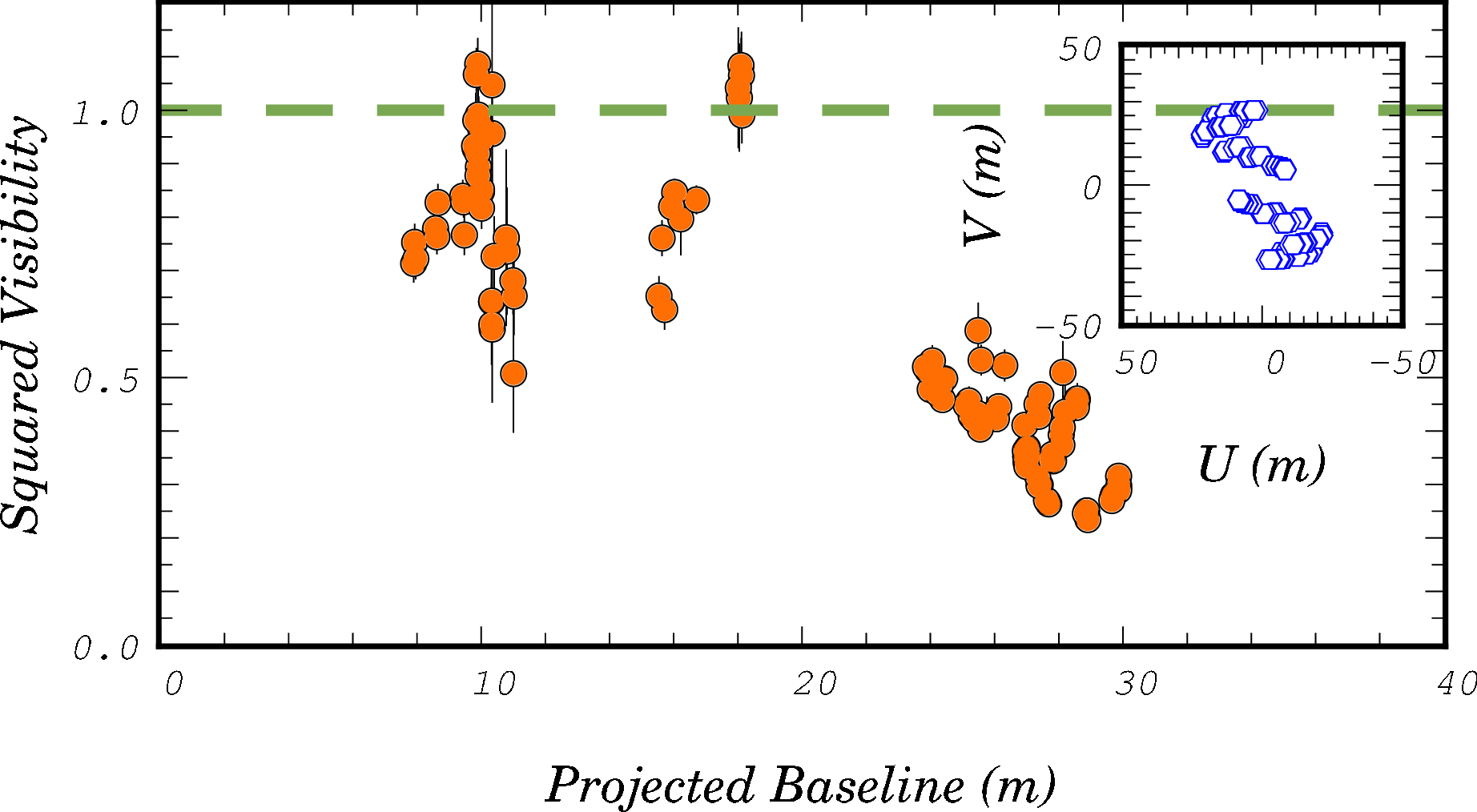}
& \includegraphics[width=0.30\textwidth]{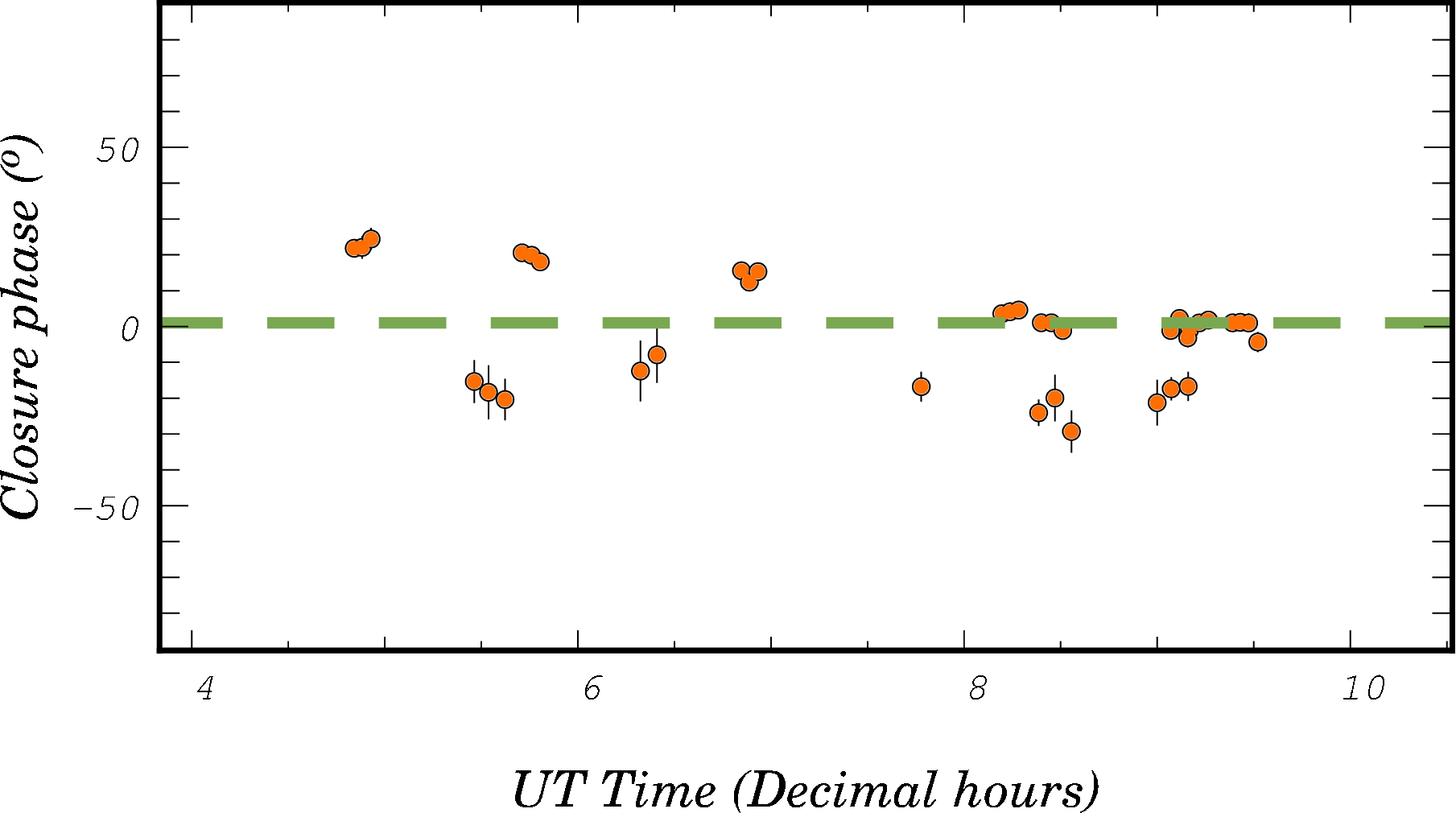}\\
\includegraphics[width=0.30\textwidth]{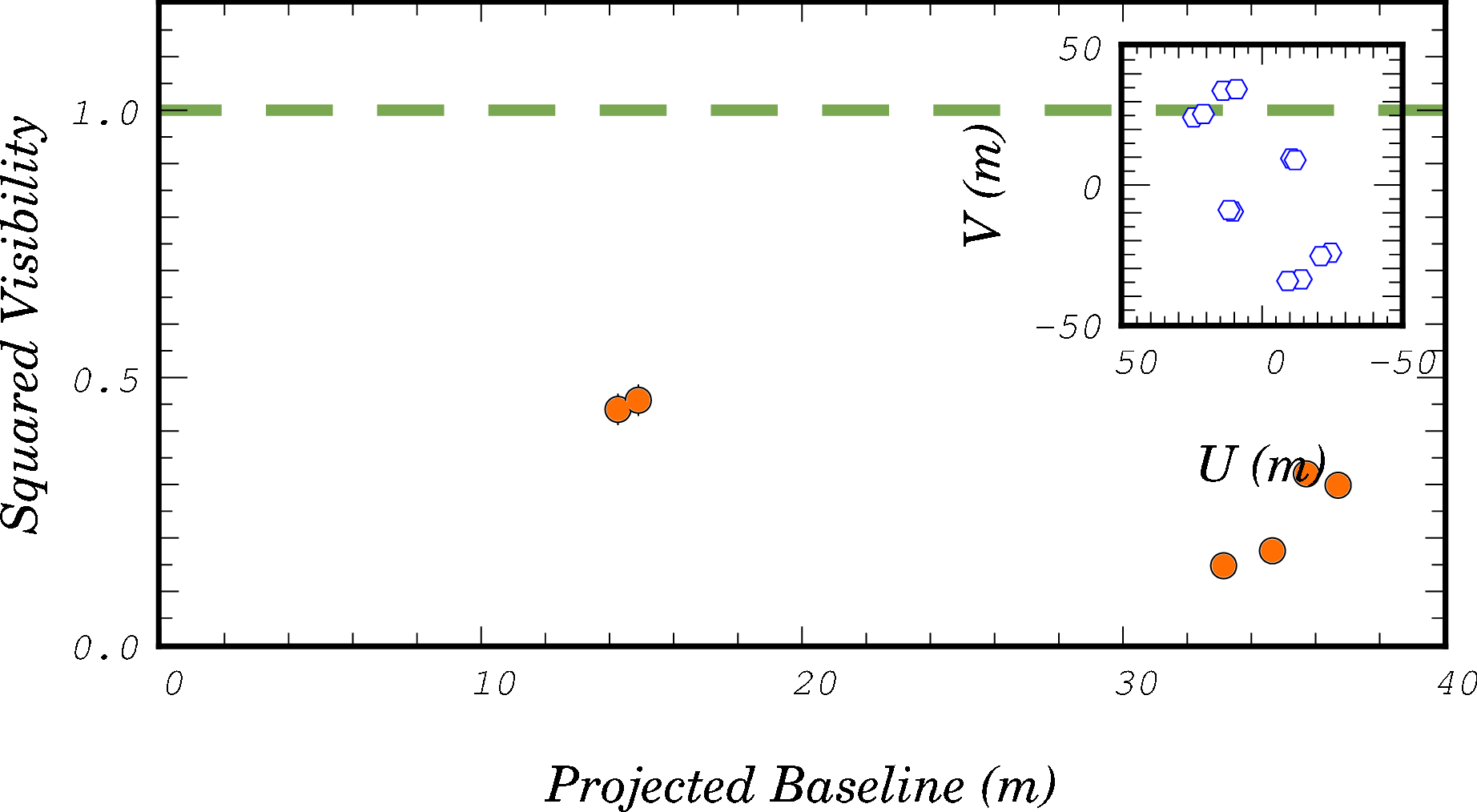}
&  \includegraphics[width=0.30\textwidth]{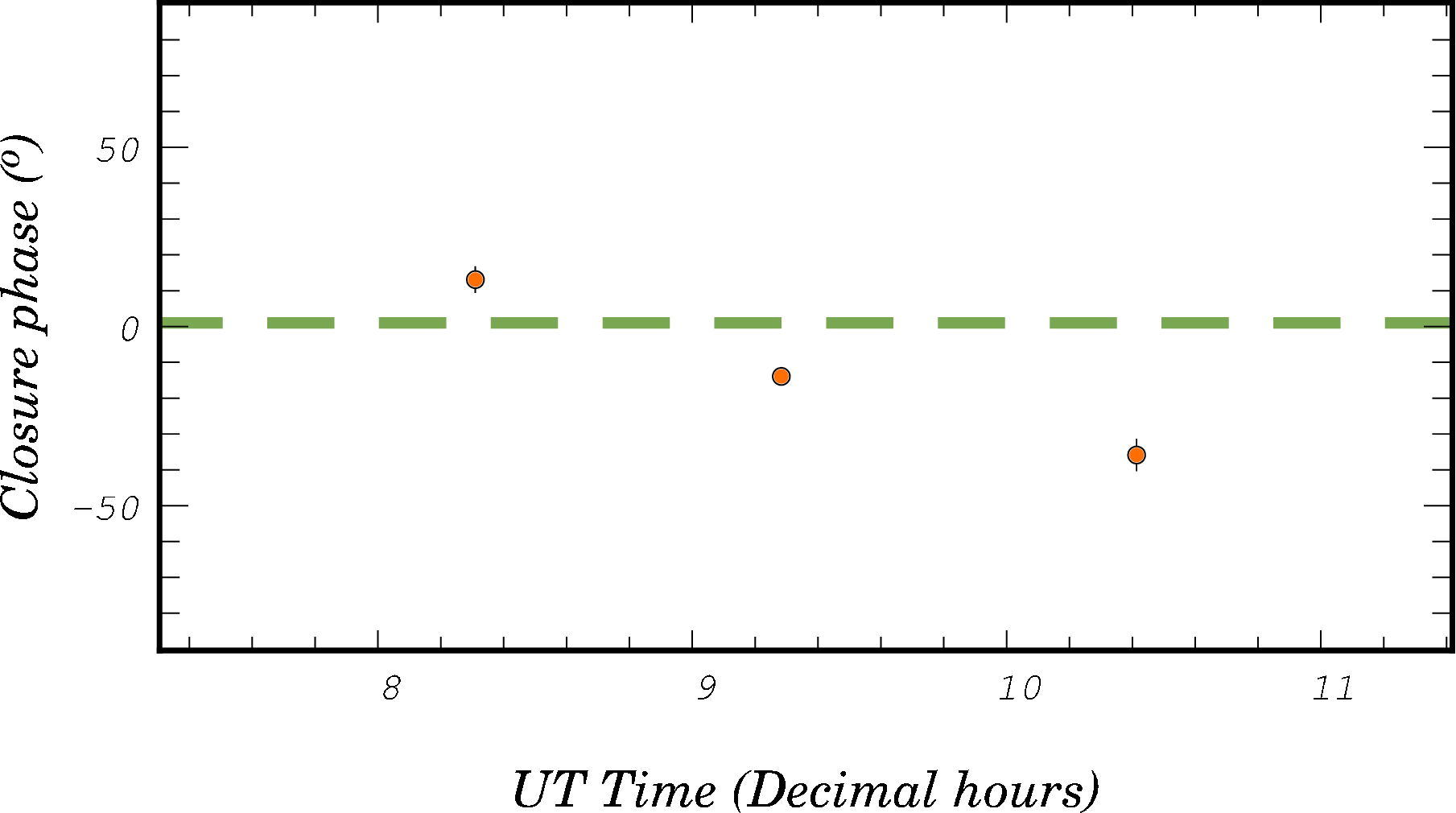}\\
\end{tabular}
  \caption{Squared visibility as a function of the projected baseline
    (left) and closure phases as a function of UT time for the epochs
    2003, 2004, and 2005 (right). 
The inset shows the corresponding UV coverage.}
  \label{fig:visib}
\end{figure*}

\begin{table*}[t]
\small
  \centering
\begin{tabular}{|c|c|c|c||c|c|c||c|}
\hline
 Epoch & $f_{B}/f_{A}$ & $\rho_{AB}$ (mas) & $\theta_{AB}
 (^\circ)$ & $f_{C}/f_{A}$ & $\rho_{AC}$ (mas) &
 $\theta_{AC}(^\circ)$ & $\chi^2$ \\
\hline 
\hline 
2003& 0.64$\pm$0.19 & 3.3$\pm$0.6 & 26$\pm$24 & 0.21$\pm$0.02
& 22.7$\pm$3.3 &247$\pm$8 & 0.6\\
\hline
2004& 0.57$\pm$0.05 & 3.4$\pm$0.8 & 189$\pm$24 & 0.24$\pm$0.02 & 18.6$\pm$1.3 &207$\pm$10 &1.6\\
\hline
2005& 0.53$\pm$0.18 & 3.4$\pm$0.6 & 17$\pm$20 & 0.27$\pm$0.1 &
18.7$\pm$1.6 &182$\pm$9 & 0.1\\
2005&\multicolumn{3}{c||}{(alternative outer solution)}& 0.27$\pm$0.05
&12.2$\pm$2.4 &152$\pm$11 & 0.1\\
\hline
2004 (from MIRA) & 0.7 & 3.5 & 171 & 0.32 & 16  & 217 & NA\\
\hline
\end{tabular}
 \caption{Triple system parameters: flux ratio, separation, position
   angle, associated errors, and reduced $\chi^2$ as obtained from the fitting procedure
   and MIRA reconstruction software.}
  \label{tab:tripleparam}
\end{table*}

IOTA was a three 45cm siderostats interferometer, located on Mt.
Hopkins Arizona and operated by Smithsonian Astrophysical Observatory
\citep{Schloerb:2006}. The three telescopes (A,B,C) were
movable along two perpendicular axes and could occupy 17 stations
offering a way to synthetize a beam with a maximum $35\times15$ m 
aperture. This corresponds to a resolution of $\approx 5\times12$mas
in the $H$-band.  Observations took place during
period 2003 November - 2005 November. Four array configurations were
used to pave the UV plane.  The observations were made in the $H$ band
using the
IONIC3 instrument \citep{Berger:2003}. IONIC3 allowed each of the three
baselines to be combined separately, which provided six
interferograms. Fringe patterns were temporally
encoded, detected, and locked with a PICNIC detector array
\citep{Pedretti:2004p5147,Pedretti:2005}.

We used the data reduction method described in
\cite{Monnier:2004p6916} to extract squared visibilities ($V^2$) and
closure phases (CP) from the interferograms. Instrumental $V^2$
and CP effects were removed by interspersing each science target
acquisition with observations of calibrators with small and known
diameters. Table \ref{tab:logs} shows a log of our observations. While
the calibrators were chosen to match the
brightness and spectral type of GW Ori as closely as possible to avoid instrument-induced
biases, we estimated that a conservative induced CP systematic error
of $\approx 0.5^\circ$ had to be added quadratically to the
statistical error computed from the set of interferograms
\citep{Monnier:2006}. This additionnal term is essentially dominated
by the chromatic response of the beam combiner.  Identically, our
night-to-night calibration accuracy follow-up led us to adopt a
conservative systematic $V^2$ calibration error of
$\sigma_{V^{2}}=0.05$ added quadratically to the statistical error on
$V2$. Our
calibrated data set resulted in the following statistically
independent measurements: 27 $V^2$ and 9 CP in 2003, 111 $V^{2}$
and 37 CP in 2004 and 6 $V^2$ and 3 CP measurements in 2005
. The calibrated data are available from the authors in the OI-FITS
format \citep{Pauls:2005p3854}.

The calibrated $V^2$ and CP data are presented in Fig.\
\ref{fig:visib}. The conclusions that can be drawn from an inspection of
these data include: 1) GW Orionis is clearly resolved at most of the
spatial scales probed by IOTA; 2) The $V^2$ and CP curves show
  an oscillation pattern typical of a multiple component system; 3) there
is a clear non-zero closure phase signal at all epochs, which hints at
non centro-symmetric emission.  The contribution of
  the circumprimary disk's H-band emission to the $V^2$ curve
  are most probably negligible because their expected angular size is
  smaller than the close binary separation, which is already
  underresolved by our observations. We note that the CB disk inner rim
  is too far away to contribute significantly to the H-band emission.  This inspection combined with the already known
binarity of GW Orionis and the evidence of a putative companion to the
spectroscopic system (\citet{Mathieu:1991} and Latham
priv. comm.) has led us to focus our analysis on the search for
sources of point-like emission in the data.

\section{Modelling and results}
\subsection{Visibility and closure phase modelling}

For the purpose of our fit we temporally binned the data by
considering three bins, i.e. 2003 (2003-11-30 to 2003-12-01), 2004
(2004-12-12 to 2004-12-20) and 2005 (2005-11-22). These three epoch
bins allowed us to neglect orbital motions within each fit (even for the
longest epoch, 2004, the eight days of observations only span 3\% of
the orbit, and therefore the orbital motion introduces negligible
errors). We first fitted simultaneously $V^2$ and CP data
with a binary model. To
achieve this, we carried-out a $\chi^2$ fine grid computation in search
for the minimum.  We exhaustively searched the parameter space of
flux ratios ($f$), component separations ($\rho$), and position angles
($\theta$).  The resulting fits at all epochs were poor and
unacceptable (reduced $\chi^2>30$), which lead us towards a triple
system model as hypothezised by \citet{Mathieu:1991}. 
All components were considered to be unresolved and the fine grid covered a
range of realistic values. The fitting procedures defined all the
above-mentioned parameters with respect to a reference source (A),
i.e. not necessarily the primary in the usual sense of it.  The
fitting procedure revealed that epoch 2005, for which we had the least
amount of data, allowed two equally plausible parameters for the
tertiary component (C), and both solutions are included.  We confirm
that the triple model improves the reduced $\chi^2$ considerably,
finding 0.6, 1.6, and 0.1 for 2003, 2004, and 2005 respectively. The
model parameters and associated errors can be found in
Table~\ref{tab:tripleparam}.  We find that the component flux ratio
remains roughly constant for all epochs, which gives confidence in our
identification of the components.  We adopt the notation that the
brightest component is referred to as ``A'', the other component in
the close pair as ``B'', and the new outer component as ``C.''  We
continue our discussion of these results in Sect.\ \ref{discussion}.

\subsection{Image reconstruction}

\begin{figure*}[t]
  \centering
\begin{tabular}{ccc}
\includegraphics[width=0.30\textwidth]{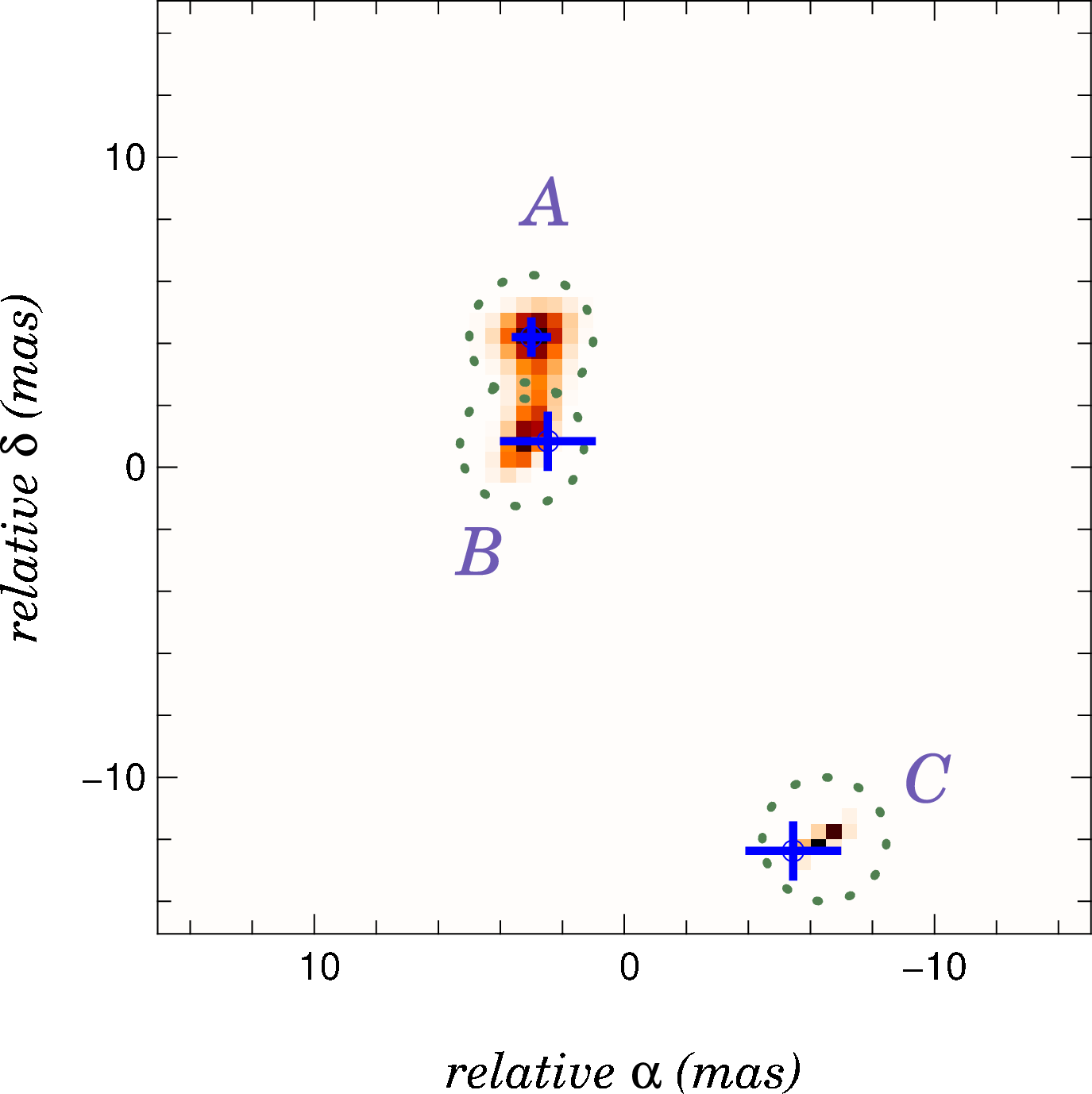}
\includegraphics[width=0.30\textwidth]{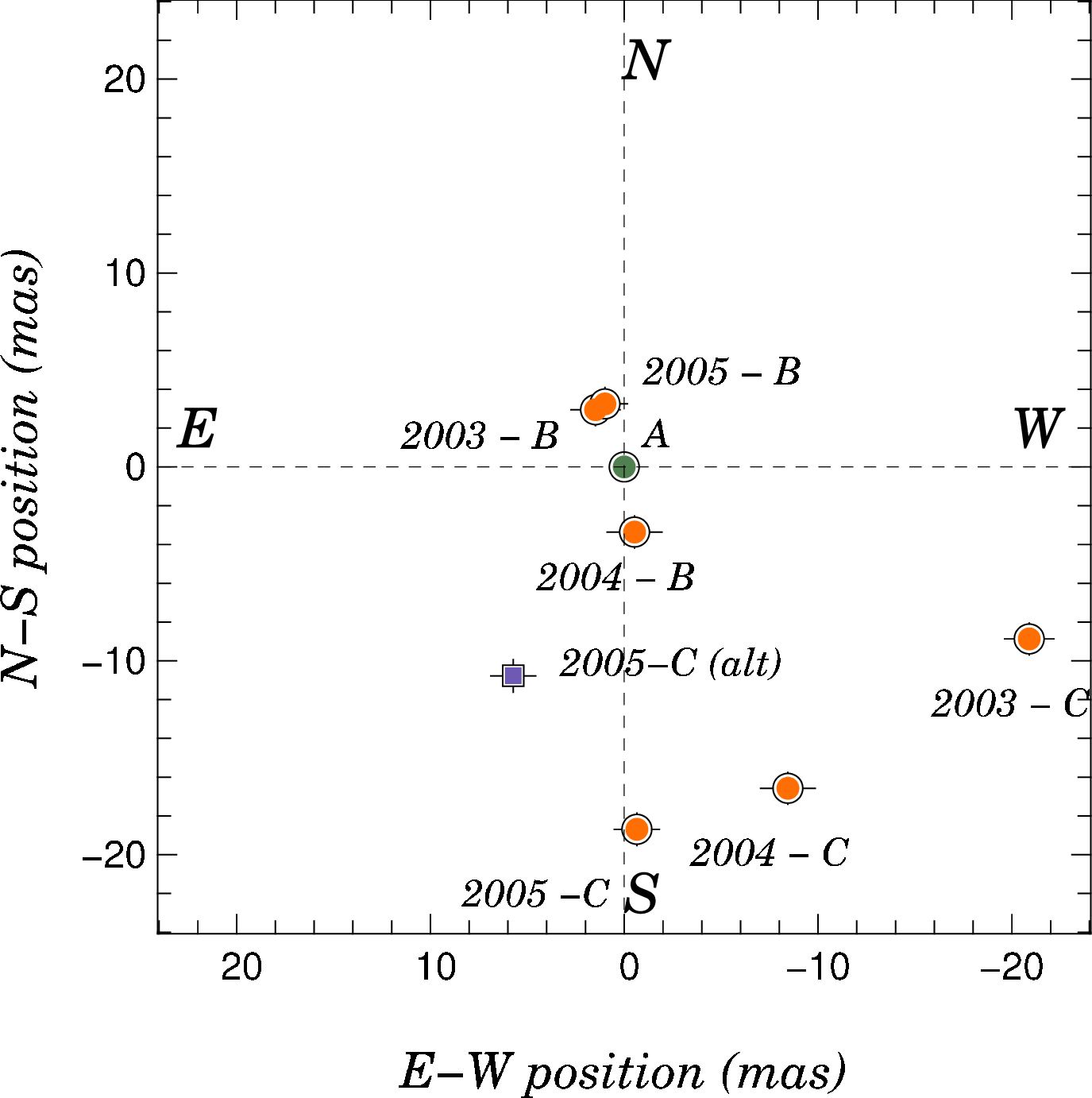}
\includegraphics[width=0.30\textwidth]{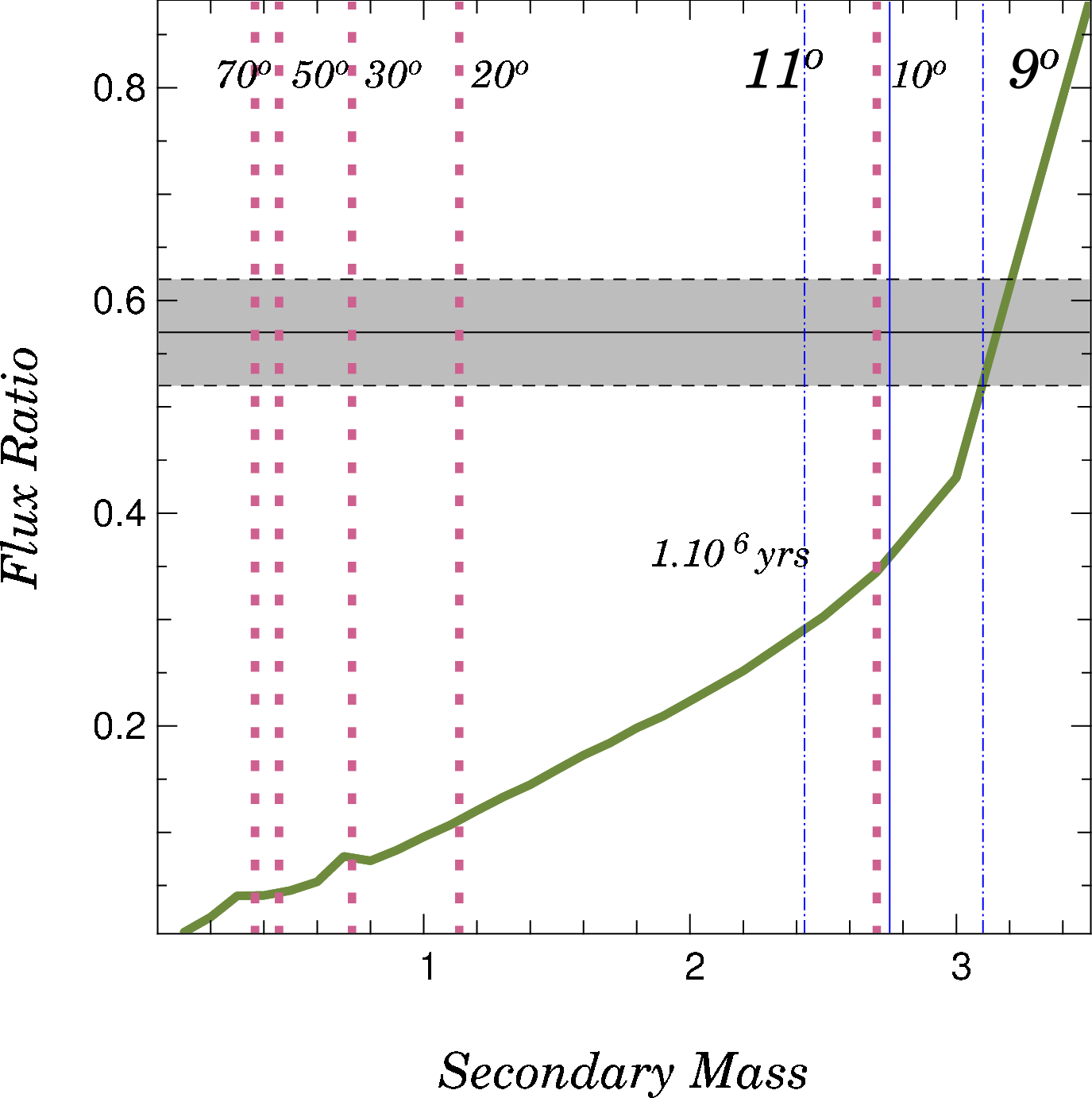}
\end{tabular}
  \caption{Left: Reconstructed image of the 2004 epoch using the MIRA software. Three components are resolved
  (encircled dots) and identified according to their brightness (A, B
  and C respectively from brightest to dimmest in H). Blue crosses indicate
the positions as obtained from model fitting. Centre: presentation of the data from
    Table~\ref{tab:tripleparam}. Square point represents the
    alternative position for epoch 2005. Right: Predicted H-band flux
    ratio as a function of the secondary mass with a primary mass of
    $3.6M_{\odot}$. Shaded region is the flux-ratio constraint from
    our observations (epoch 2004). Vertical dashed lines provide the
    system inclination. Vertical blue lines are the secondary mass
    estimate corresponding to distance $d\approx414\pm7$pc. \label{fig:tripleparam}}
\end{figure*}

We used the MIRA software \citep{Thiebaut:2008}
to independently reconstruct an image out of the V2 and
CP data. Only epoch 2004 has sufficient data for such an
attempt. This model-independent image (see Fig.\ \ref{fig:tripleparam}) is
found to support our choice of the triple system discussed in the
previous section. Furthermore, even though the procedures are
different, the locations of the three components from the image are found
to be consistent with the best-model predictions (see Table \ref{tab:tripleparam} for a comparison).

\section{Discussion}
\label{discussion}

The system is resolved into three components, the spectroscopic pair
(A \& B) plus a new outer component (C).
 The inner pair shows a
position reversal from 2003 to 2004 and also 2004 to 2005.  This is
expected since the orbital period has been measured to be $\approx 242
$ days --
furthermore, our observations happen to coincide with times when
the components are approximately in the plane of the sky (circular orbit phases of
$\sim$0.0, 0.6, 0.0 in 2003, 2004, 2005 respectively). Unfortunately,
this timing means that our visual orbit cannot constrain the orbital
inclination.  On a positive side, this means that our measured
separation is close to the actual semi-major axis of the orbit,
1.35$\pm$0.10~AU assuming a distance of 400pc and knowing that the
eccentricity of the orbit is $\approx 0$ \citep{Mathieu:1991}. This
corresponds to a system mass of 5.6$^{+1.4}_{-1.1}$~M$_{\odot}$, somewhat
higher than the previous estimate of 3.0~M$_{\odot}$\footnote{We note
  that introducing the uncertainty on the distance (i.e $400^{+21}_{-32}$pc)
  causes an error of $\approx$ 20\% in the mass estimate which
  is compatible with this remark.} \citep{Mathieu:1991}.

Perhaps the most surprising aspect of this inner pair is that the
H-band flux ratio is only about a factor of $\sim$2.  Previous work
\citep[e.g.,][]{Mathieu:1991} had estimated
that the system was composed by two stars of respectively
2.5$M_{\odot}$ and 0.5$M_{\odot}$ and concluded that the contribution
of the secondary to the near-infrared excess was negligible.

To investigate the link between the flux ratios and the masses we
  used the pre-main-sequence stars theoretical evolutionary
  models computed by \citet{siess:2000}, which were shown to be
  reliable enough to estimate masses with a 10\% to 20\% precision
  \citep{Hillenbrand:2004}. We considered, as in
  \citet{Mathieu:1991}, that the effective temperature of the primary
  was $T_{\rm{eff}}\approx 5700K$ and the age $\approx 10^6 yrs$. We
  interpolate from the $10^6$yrs isochrones that this
  corresponds to a primary mass of $\approx 3.6 M_{\odot}$. The rightmost
  plot in figure \ref{fig:tripleparam} shows the expected flux ratio
  from the \citet{siess:2000} computation as a function of secondary
  mass. The straight line materializes the observed flux ratio (i.e
  $0.57\pm 0.05$, 2004).  Additionally, we use the mass ratio and mass
  function determined from radial velocity to compute the requested
  orbital inclination as in \cite{Mathieu:1991}.

If we take the most recent and precise determination of the distance
by \cite{Menten:2007}, i.e $414\pm 7$ pc, we can extract from figure
\ref{fig:tripleparam} the following information:

\begin{itemize}
\item Our observation can be explained by a binary system with a mass ratio close to unity
  at low inclination and no excess from circumstellar emission. 
Indeed, taking the uppermost distance estimate of 421 pc the observed flux ratio in the H band can be reproduced by a
  system seen at low inclination ($i\approx9^\circ$) with a secondary
  mass of $\approx 3.1 M_{\odot}$, i.e a mass-ratio of
  the order of $0.86$; 
\item With decreasing distance the observed flux ratio requires the
  addition of H-band excess to the secondary and a slight increase in
  inclination. For example, if we take the lowest distance boundary of 407 pc, the
expected  secondary-to-primary ratio is $\approx 0.29$, which requires
a secondary circumstellar excess  of roughly the same value $\approx
0.28$. The requested inclination is then $\approx 11^{\circ}$
\end{itemize}

Although this analysis is subject to important uncertainties, including
the primary spectral type estimate, we tested that our conclusions hold
at a qualitative level, i.e that \emph{the GW Ori inner system is
  probably seen at a much lower inclination (i.e $\approx 10^{\circ}$) with
  more circumsecondary H
  band emission than previously thought \citep{Mathieu:1991}.}
Reconciling our observation with the observed eclipses by
\citet{Shevchenko:1998} is a puzzle we cannot solve.

Another important discovery of our work is the existence of source C.
We detected C in each epoch and found clear evidence for motion
with respect to the inner pair AB.  We compared our motion to the
preliminary 3850~day orbit recently inferred from long-term radial
velocity monitoring (Latham, priv. comm.) but did not find a
solution that could simultaneously fit our motions and the
spectroscopic orbital elements.  If the 20~mas separation of AC is the
true semi-major axis, then we would expect an orbital period of
$\sim$3600~days, which is quite close to the period seen in the RV residuals.
It is far from clear if this hierarchical triple is stable.
Additional
interferometric monitoring of the inner pair will allow the
inclination to be measured directly and the matter will be settled
quantitatively in the near future.

\section{Conclusion}

We have presented the first direct detection of the CTTS GW Orionis as
a triple system: {\bf both the secondary and tertiary companion are
spatially resolved}. Our results show that the inner stars have H-band fluxes
within a factor of 2 of each other, which suggests that the exact status
of the pair should be revisited. The outer companion is clearly
detectable, consistent with a roughly 3500~day orbital period, in line
with hints from radial velocity residuals.  GW Orionis offers a unique
laboratory to study the dynamical evolution of a close, young and
(perhaps) hierarchical triple system and its interaction with a
massive disk and envelope.  Rapid progress in determining the full
orbital elements of the three components is possible with current
interferometers in the coming years.

\begin{acknowledgements}
  This work was supported by ASHRA, PNPS/INSU,
and  Michelson fellowship program.  This research has made use of the
  SIMBAD database, operated at CDS, Strasbourg, France and getCal
  software from the NASA Exoplanet Science Institute, Caltech. Bibliographic references were provided
  by the SAO/NASA Astrophysics Data System.  IONIC-3 has been
  developed by LAOG and CEA-LETI, and funded by the CNRS and CNES. We
  thank Russel White, Lee Hartmann and S. Meimon for useful
  discussions.
\end{acknowledgements}

\bibliographystyle{aa}
\bibliography{gworiI,biblio}

\end{document}